\documentclass[twocolumn,showpacs,preprintnumbers,amsmath,amssymb]{revtex4}

\usepackage{dcolumn}
\usepackage{bm}

\usepackage{latexsym}
\usepackage{amsfonts}
\usepackage[dvips]{graphicx}
\usepackage[usenames]{color}
\usepackage{makeidx}

\newcommand{\ket}{\rangle }
\newcommand{\bra}{\langle }

\begin{document}


\title{Breakdown of a Mott insulator --- 
non-adiabatic tunneling mechanism}
\author{Takashi Oka, Ryotaro Arita, Hideo Aoki}
\address{Department of Physics, University of Tokyo, Hongo, Tokyo 113-0033, 
Japan}

\date{\today}
\begin{abstract}
\noindent 

Time-dependent nonequilibrium properties of a strongly correlated 
electron system driven by large electric fields is obtained 
by means of solving the time-dependent Schr\"odinger equation for 
the many-body wave function numerically in one dimension. 
While the insulator-to-metal transition depends on the electric 
field and the interaction, the metallization is found to be described 
in terms of an universal Landau-Zener quantum tunneling among the many-body levels.
These processes induces current oscillation for small systems, 
while give rise to finite resistivity through dissipation 
for larger systems/on longer time scales.  

\end{abstract}

\maketitle

{\it Introduction} Nonlinear responses and time-dependent phenomena are an 
open frontier in the physics of strongly correlated systems.  
Namely, while the physics of electron correlation has been 
intensively studied in the last decades in the context of the 
high-Tc cuprates and related systems, 
the efforts were concentrated on equilibrium or linear-response properties. 
Properties beyond the linear response or
time-dependent properties have been much discussed\cite{Noneq}, 
but remains a challenging problem, 
which is, theoretically, due to the difficulty in
dealing simultaneously with the many-body effect (correlation) and
finite current (non-equilibrium), which are both non-perturbative physics.
So we definitely require a clear picture for 
this fundamental problem, especially for 
low-dimensional materials.  

In this Letter we consider the Hubbard model 
under constant driving forces in one dimension.  
The ground state of the half-filled Hubbard system 
is a Mott insulator\cite{Lieb} for arbitrary strengths of 
the electron-electron repulsion $U>0$, while the state is metallic 
when the band filling is shifted away 
from the half-filling by doping carriers. 
So the question we pursuit here is: 
what will happen if we destroy the Mott insulator by applying strong 
electric field, instead of doping? 

There are existing theoretical approaches that employ the Bethe-ansatz method, 
and the closing of the Mott gap has been discussed\cite{Fukui,Deguchi:1998GC}.  
In these studies, however, electric fields are not actually applied, 
but the left-going and right-going hopping terms are 
made different instead.  It is rather difficult to relate this 
artificial, and non-Hermitian, model with a system in an electric field.  
That is why we have here opted for actually applying electric 
fields for the first time to keep track of 
the time evolution of the many-body wave functions and levels. 
For that we make use of a numerical integration of the 
time-dependent Schr\"{o}dinger's equation, where 
we have also looked at excited states, which should be essential 
in examining non-equilibrium phenomena.  

We have found that the electric field, if strong enough, 
breaks the Mott-insulator phase.  
While the critical field strength required for the 
breakdown of the Mott insulator depends sensitively on the 
magnitude of the electron-electron interaction, 
we propose here that the mechanism for the metallization can be viewed as 
the non-adiabatic tunneling between the many-body levels.  
We have verified this by confirming numerically that 
a universal Landau-Zener quantum tunneling governs the 
nonlinear conduction (e.g., the $I-V$ characteristics representing the 
insulator-to-metal transition), where the fit is surprisingly good 
although the Landau-Zener formalism is originally intended for 
one-body problems while the problem at hand, being many-body, 
involves the Hilbert space with huge dimensions.  

Just after the metallization, we have a self-induced current 
oscillation.  While this should be realistic for mesoscopic 
systems, for large systems or on a longer time scale, 
a novel, ``Ohmic" conduction is found to 
result.  This occurs despite the absence of 
disorder and the heatbath degrees of freedom, but 
a series of non-adiabatic tunneling among many-body states 
are responsible for the dissipation effect.  
Indeed, the  breakdown of 
a one-dimensional (1D) \cite{tag} as well as 
two-dimensional (2D) \cite{yama} Mott insulator 
has been experimentally studied, and, 
among other interesting phenomena 
including spontaneous density-pattern formations,\cite{yama,tag,Kuma} 
a seemingly Ohmic conduction was found 
for a rather wide range of external electric field, until the 
Ohmic conduction is eventually broken.  

\begin{figure}[htbp]
\centering 
\includegraphics[width=6.5cm]{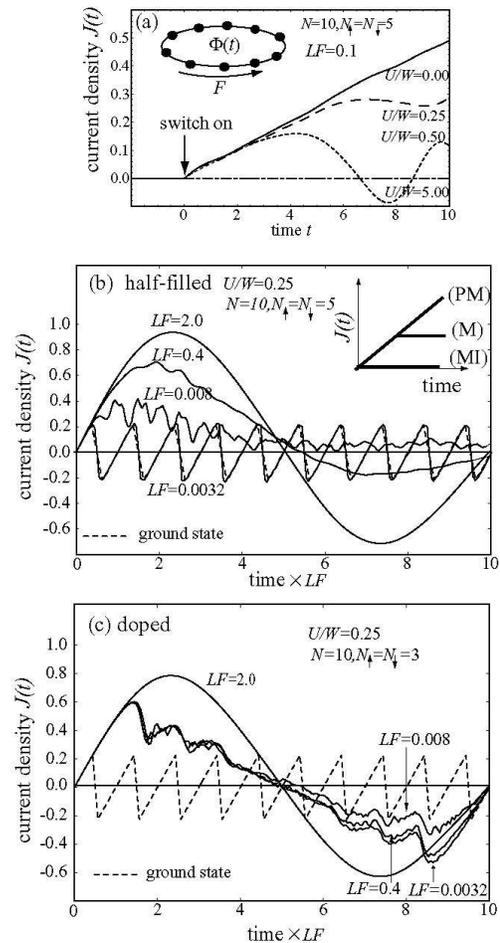}
\caption{(a) Time evolution of 
the current, $J(t)$, for a half-filled, 10-site 
Hubbard model for 
various strengths of the correlation, $0\leq U/W\leq 5$, 
for a fixed electric field $F=1/10L$.   
Time is measured in units of $\tau_W\equiv 4\hbar/W$, 
$LF$ in $W/(4\hbar)$, and $J(t)$ in 
$1/\tau_W$. The range of the time in this panel corresponds to a range of 
the AB-flux $0\le\Phi/\Phi_0\le 1$.  
The inset schematically depicts the sample geometry, where 
an AB flux, $\Phi(t) = LFt$, increasing linearly with time induces 
an electric force through Faraday's law.
(b) A longer-time behavior of the current 
when $F$ is varied with a fixed 
$U/W=0.25$, again for the half-filled case. 
Here the horizontal axis is $LF t = \Phi(t)/\Phi_0$ and 
over $0\le \Phi/\Phi_0\le N (=10$ here).  
The inset schematically shows three kinds of behavior (MI: Mott 
insulator, M: metal, PM: perfect metal).
(c) A plot similar to (b) for a non-half-filled case 
($N_\uparrow=N_\downarrow=3<N/2=5$).
}
\label{currents}
\end{figure}

{\it Formulation} If one wants to apply a finite electric field to a system, 
one problem is how to treat the electrodes.  
To get rid of ambiguities arising from this, we have opted here for a 
periodic system (a ring in 1D), where 
the electric field $F$ is applied via a time-dependent 
AB flux $\Phi(t) = eLFt$ piercing the ring (inset of Fig.\ref{currents}(a); 
$L$: sample length). 
This will lead to a circular electromotive force due to Faraday's law.   
The flux makes the hopping integral in a tight-binding 
model complex, where the Hamiltonian is
\begin{eqnarray}
H(t)&=&-\frac{W}{4}\sum_{i,\sigma}\left(\;e^{2\pi i\Phi(t)/N}c_{i+1\sigma}^\dagger c_{i\sigma}+{\rm h.c.}\;\right) \nonumber\\
&+&U\sum_{i}n_{i\uparrow}n_{i\downarrow}. 
\label{ham}
\end{eqnarray}
Here $W$ is the bandwidth, $U$ the electron-electron repulsion, 
$N=L/a$ the total number of sites with $a$ being the lattice constant.
Hereafter we take the unit in which $e=a=h=1$.
The electric field in this formalism 
amounts to a time-dependent deformation of the Hamiltonian.


We have then to solve the time-dependent Schr\"odinger equation, 
$i\frac{d}{dt}|\Psi (t)\ket=H(t)|\Psi (t)\ket$,
which governs time evolution of the quantum system at absolute zero 
temperature, starting from  
the ground state of the Hamiltonian at $t=0$, $|\Psi (t=0)\ket\equiv 
|\Psi_0\ket$, which is obtained with the Lanczos method.  
The time-integration of the state vector, which, being many-body, 
has a huge dimension, requires a reliable algorithm.  
So we adopt here Cranck-Nicholson's method\cite{CranckNicholson} that gurantees the unitary 
time evolution, where the time evolution is put in a form,
$
|\Psi (t+\Delta t)\ket=e^{ -i\int_t^{t+\Delta t} H(t)}\; dt\;|\Psi (t)\ket\simeq
[1-i\Delta t/2H(t+\Delta t/2)]/[1+i\Delta t/2H(t+\Delta t/2)]\;|\Psi (t)\ket,
$
which is unitary by definition.  
Here the time step is taken to be small enough ($dt=1.0\times 10^{-2}$ with the time in units of $4\hbar/W$ hereafter) 
to ensure convergence for $N\leq 10$ site systems, 
for which the dimension of the Hamiltonian is $\sim 10^4$.  
We have concentrated on the total $S^z=0$ subspace with 
$N_\uparrow=N_\downarrow=N/2$.  

{\it Result} 
We first plot in Fig.\ref{currents}(a) 
the result for the expectation value of the 
current density averaged over the sites, 
\begin{equation}
J=-\frac{W}{4\;N}\sum_{i,\sigma}\left( ie^{2\pi i\Phi/N}\;c_{i+1\sigma}^\dagger c_{i\sigma} + {\rm h.c.}
\right). 
\end{equation}
The behavior of $J(t)$ 
for various values of $U$ with a fixed value of the electric field $F$ 
is seen to fall upon three regimes when $U$ is varied: 
A perfect metallic behavior ($J(t)\propto t$) 
when the electrons are free ($U/W=0$), 
while when the interaction is strong enough 
($U/W\gg 1$) the current has a zero expectation value. 
For an intermediate regime of $U/W$ we have finite $J$'s 
with some oscillations in the current for finite systems.  
By contrast, a non-half-filled system has the time evolution 
distinct from the ground-state behavior.  
The difference has its root in the spectral property 
as will be discussed later.  

Now, Fig.\ref{currents}(b) plots 
the time evolution of the current 
when the electric field $F$ is varied with a fixed 
value of $U/W$, again for the half-filled case.  
The result may be summarized as follows: 
\begin{description}
	\item[(i)] Small $F$ regime (Mott insulator)\\
A drastic difference between the half-filled and doped systems appears 
for small $F$.  When half-filled, $J(t)$ in the limit of 
$F\rightarrow 0$ smoothly approaches 
the $\langle J(t)\rangle=0$ 
behavior of the ground state (Mott insulator).  Here 
$\langle J(t)\rangle$ is the time-averaged current.  
On top of the $\langle J(t)\rangle=0$ an oscillatory behavior 
with the period of $\Phi_0 (\equiv e/h=1$: 
flux quantum) is seen, which 
is nothing but the AB oscillation (a saw-tooth, due to 
a symmetry about $\Phi=\Phi_0/2$).  

\item[(ii)] Moderate $F$ regime (metal)\\
In this regime, the current in the half-filled case 
shows an oscillatory behavior 
(see typically the $LF=0.008$ data in 
Fig.\ref{currents}(b)).

 	\item[(iii)] Large $F$ regime (perfect metal)\\
When the electric field $F$ becomes large enough, 
the system becomes a metal, in which $\langle J(t)\rangle \propto 
t$ for $\Delta\Phi<\Phi_0 N/4$.  A further oscillation in $J(t)$, with a 
long period ($\Delta\Phi=\Phi_0 N$), is seen, which we will discuss later.  

\end{description}

\begin{figure}[ht]
\centering 
\includegraphics[width=6.5cm]{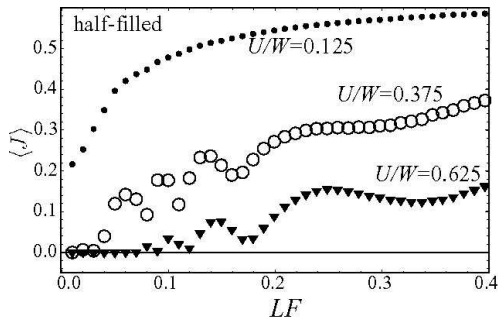}
\caption{$I$-$V$ characteristics for various 
values of $U/W$ for the half-filled Hubbard model with $N=6$. 
}
\label{IV}
\end{figure}

The $F$-dependence of $\langle J\rangle$ 
are displayed in Fig.\ref{IV} as 
the $I$-$V$ characteristics for various values of $U$.  
Here the time average $\bra J\ket = \int_0^{\Delta t}\bra J(t)\ket dt/\Delta t\;(LF\Delta t=N\Phi_0/4)$ of the current
density $J(t)$ is taken over 
one forth of the extended AB period ($0\leq \Phi \leq N\Phi_0/4$), since 
we are interested in the rise in the current which should represent the 
behavior in the thermodynamic limit.  
We can see that $\langle J \rangle$ becomes nonzero rather 
abruptly at the metallization, where the threshold electric 
field increases with $U/W$, thereby the $F$-
dependence becoming weaker.  Just after the metallization 
some oscillation (in the $F$-
dependence this time) is seen for finite systems.

\begin{figure}[ht]
\centering 
\includegraphics[width=8.5cm]{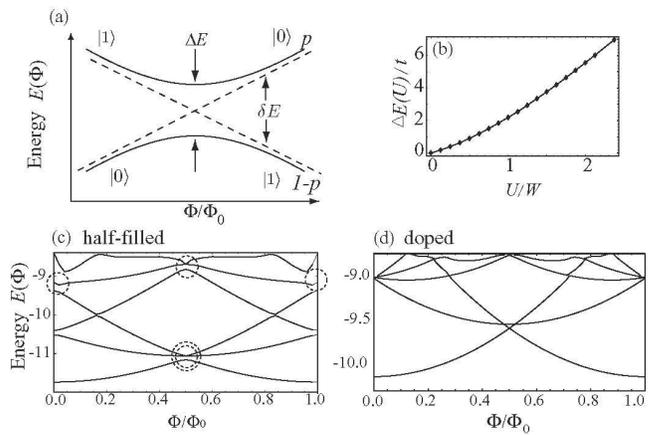}
\caption{
(a) Landau-Zener process around a level anticrossing, 
where $\Delta E$ denotes the gap, 
$p$ the transition probability, and $\delta E$ the 
difference between the ``unperturbed", crossing energy levels 
(dashed lines).  
(b) $U$-dependence of the many-body 
gap $\Delta E$ marked with a double circle in (c). 
(c) The low-lying levels versus $\Phi$ 
for the Hamiltonian eq.(\ref{ham}) in the 
half-filled case ($N=10$ with $N_\uparrow=N_\downarrow=5$) for $U/W=0.125$. 
Level repulsions due to the interaction $U$ are encircled.  
(d) Similar plot for a doped case 
($N=10$ with $N_\uparrow=N_\downarrow=3$).
}
\label{spectrum}
\end{figure}

{\it Non-adiabatic tunneling} 
In order to understand the physics underlying these time evolutions and 
their $F$ dependence, let us here evoke the notion of non-adiabatic 
tunneling, originally conceived for one-body problems 
by Landau, Zener, and St\H{u}ckelberg (LZS)\cite{Landau,Zener,St}. 
The LZS theory considers the quantum tunneling beyond the adiabatic 
approximation.  Namely, when a parameter defining the Hamiltonian is varied 
infinitely slowly (adiabatically), we can just plot a set 
of energy levels against the parameter, which in general contain 
level anticrossings, since the two levels should repel with each 
other unless they are allowed to cross due to e.g. symmetry reasons.  
An initial state that start from 
one of the lines should evolve with time sticking to that line 
(adiabatic theorem, corresponding to $p=0$ in Fig.\ref{spectrum}(a)).   
When the parameter is varied with a finite velocity, 
the state can make a transition across the level anticrossing with 
a finite probability $p (\neq 0$ in Fig.\ref{spectrum}(a)), where 
the transition is caused by a quantum mechanical tunneling across 
the gap $\Delta E$. 
The transition probability $p$ depends in the LZS theory
on the speed ($\delta \dot{E}$) the two energy level approach as 
\begin{equation}
p=\exp\left(-2\pi \;\frac{(\Delta E)^2}{\delta \dot{E}} \right)
=\exp\left(-2\pi\;\frac{(\Delta E)^2}{ \frac{d\delta E}{d\Phi}LF }\right). \label{lzs}
\end{equation}
Here $\delta E$ is the difference between the ``unperturbed", 
crossing energy levels (dashed lines in Fig.\ref{spectrum}(a)) and 
$\delta \dot{E} \equiv d\delta E/dt = (d\delta E/d\Phi)\dot{\Phi}$ 
with $LF=d\Phi/dt$. 
We can immediately see that the process is 
non-perturbative in $F$, since $p$ is singular in $F$.

\begin{figure}[hbt]
\centering 
\includegraphics[width=7cm]{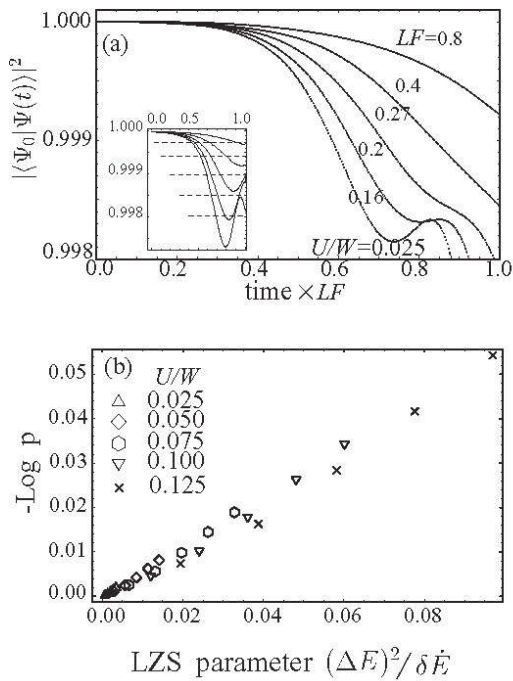}
\caption{(a) Time-evolution of the 
weight of the ground state, $|\bra \Phi_0|\Phi(t)\ket |^2$, 
calculated for the half-filled Hubbard model 
($N=10, N_\uparrow=N_\downarrow=5$) for various values of 
$F$ with $U/W=0.025$. Inset shows the 
solutions\cite{asymptotic} of the LZS equation with its asymptotic values indicated.
(b)Transition probability $p$ (decrease in $|\bra \Phi_0|\Phi(t)\ket|^2$) 
plotted against the LZS parameter $\frac{(\Delta E)^2}{\delta \dot{E}}\;(\delta\dot{E}=\frac{d\delta E}{d\Phi} LF)$ 
for various values of $F$ and $U/W$.}
\label{lz}
\end{figure}

Although the original LZS theory is devised for one-body systems, 
there is no reason why we cannot apply it to many-body systems, 
as demonstrated for a spin system by Miyashita et al\cite{raedt,Miya}. 
So here we apply the concept to the Hubbard model, which is, 
to our knowledge, the first time the LZS theory is
applied to interacting electron systems.
In order to check the validity of the LZS picture, 
we have first calculated the transition probability $p$.  
The level anticrossing we focus on is the first one 
encountered by the ground state at $\Phi=\Phi_0/2$ 
in the level flow 
(marked with a double circle in Fig.\ref{spectrum}(c)). 
In Fig.\ref{lz}(a) we plot $|\bra \Psi_0|\Psi(t)\ket|^2$, 
the weight of the ground state around the level anticrossing, 
for various values of $U$ and $F$.   
From this we have obtained the transition probability $p$ 
as the asymptotic value of $|\bra \Psi_0|\Psi(t)\ket|^2$\cite{asymptotic}.
Figure \ref{lz}(b) plots -Log $p$ 
as a function of the LZS parameter, 
$(\Delta E)^2/(\delta\dot{E})$, 
with $\Delta E$ now defined as the interaction-originated one. 
We can see a remarkably accurate linear dependence 
on the LZS parameter, which 
clearly indicates that the LZS theory is applicable to 
the many-body system we have at hand.
So the field $F$ and the interaction $U$ 
enter into the problem only as a combination $(\Delta E(U))^2/(\frac{d\delta E}{d\Phi} LF)$
(where $\Delta E(U)$ is an increasing function of $U$; 
see Fig.\ref{spectrum}(b)).    

For large enough electric field $F$ the nonadiabatic 
tunneling is so effective that the state goes 
straight across each crossing with probability close to unity, 
so we end up with a long period ($\Delta\Phi=\Phi_0 N$), 
an analogue of
the ``extended AB period" discussed for electron systems by 
Kusakabe and two of the present authors\cite{Kabe,Ari}, 
a notion originally proposed for a spin (Heisenberg) system 
by Sutherland \cite{Sut}.
  In the thermodynamic limit ($N \rightarrow \infty$) 
the period becomes $\infty$.

\begin{figure}[t]
\centering 
\includegraphics[width= 7cm]{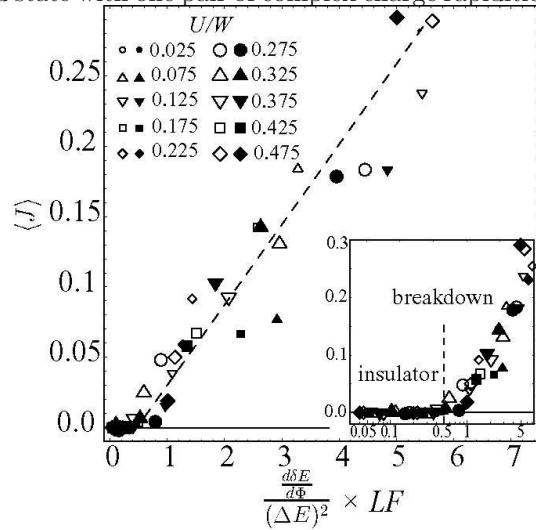}
\caption{Time-averaged current plotted against the 
inverse LZS parameter 
$(d\delta E/d\Phi)/(\Delta E)^2\times LF$
for various values of $U/W$ and $F$ in the half-filled 
Hubbard model with $N=6$ (open symbols) and $N=8$ (solid ones).  
Inset is a blowup of the same data for small 
$F$ on a log-scale. }
\label{IVlz}
\end{figure}

Encouraged by this, we have then re-plotted 
the time-averaged current, $\langle J \rangle$, against the inverse of the LZS parameter
$\frac{d\delta E}{d\Phi}/(\Delta E)^2\times LF$, in Fig.\ref{IVlz}.  
Dramatically, all of the curves, which appeared quite different 
for different values of $U/W$ 
in the raw $I$-$V$ characteristics (Fig.\ref{IV}), 
fall on a single, universal curve within a reasonable error when 
plotted against the LZS parameter.  
Specifically, a threshold between the insulating behavior 
and the dissipative metallic one 
is clearly seen at around 
$F\sim 0.5\times (\Delta E)^2/\left(\frac{d\delta E}{d\Phi} L\right)$.

After the metallization the current is seen to behave 
roughly linearly with $F$.  
This seemingly ``Ohmic" behavior is far from trivial 
but rather surprising, since we have a many-body but clean system. 
This implies that (i) after many level crossings 
the system reaches a steady state, and (ii) the many-body gaps 
at these level crossings have similar magnitudes as the 
first one ($\Delta E$ above).  
We can in fact recall that the 
non-adiabatic tunneling is a quantum version of dissipation, where 
different quantum states become mixed after level anticrossing.
To get an insight into the quantum dissipation 
let us examine the nature of the many-body gap $\Delta E(U)$ 
in the half-filled Hubbard model, which is a charge gap 
characteristic to 
the half-filled Hubbard model and vanishes when doped 
(Fig.\ref{spectrum}(c)). 
The total momenta of the anticrossing states 
(the ground state and the excited state with one pair of complex charge rapidities\cite{Woy}) 
differ by $2k_F$ where $k_F$ is the Fermi wavenumber 
(for small $U$; replace $k_F$ with the quasimomentum
for general $U$).  
At half-filling, these Umklapp processes take place 
and the momentum of the many-body state is dissipated.  
In other words, the role of heatbath degrees of freedom\cite{CaLe} 
is played by the many-body system itself.  
Let us add that the current 
oscillation (typically seen for $LF=0.008$ in Fig.\ref{currents}(b)) is 
due to the kicks from Umklapp processes and should be 
observed in small systems with strong electron correlation. 
The threshold electric force, 
$F \simeq 0.5 (\Delta E)^2/[(d\delta E/d\Phi)L]$, 
has a similar order of magnitude as the critical field observed 
in the experiment\cite{tag}.  
The threshold electric force, 
$F \sim (\Delta E)^2/[(d\delta E/d\Phi)L]$, should approach 
an asymptotic value in the thermodynamic limit, 
since $(d\delta E/d\Phi)|_{\Phi=\Phi_0/2} 
(\sim (d^2E/d\Phi^2)|_{\Phi=0} \sim$ Drude weight) $\sim 1/L$ 
should cancel the $L$ in the denominator.
Exactly how the thermodynamic limit is reached and 
how the seemingly Ohmic conduction occurs 
is an important future problem.  

We thank Koichi Kusakabe for discussions, and 
TO acknowledges Naoki Watanabe for a help on 
the time-dependent numerical simulation.



\begin{references}
\bibitem{asymptotic} We have obtained the asymptotic value of
$|\bra \Phi_0|\Phi (t)\ket |^2$ by fitting the data obtained in
Fig.\ref{lz} (a) to the exact solution\cite{Zener} 
$1-x|D_{-n-1}(z)|^2$ with $x$ as the fitting
parameter, where $D_n(z)$ is the Weber function.
\end{references}
\end{document}